\documentclass[aps, pra, twocolumn, tightenlines, footinbib]{revtex4-2}
\usepackage{amsmath}
\usepackage{amssymb}
\usepackage{mathrsfs}
\usepackage{graphicx}
\usepackage[colorlinks=true, letterpaper=true, pdfstartview=FitV, linkcolor=blue, citecolor=blue, urlcolor=blue]{hyperref}

\begin{document}

\title{Anti-parity-time symmetry hidden in a damping linear resonator}

\author{Xun-Wei Xu}
\email{Corresponding author: xwxu@hunnu.edu.cn}
\affiliation{Key Laboratory of Low-Dimensional Quantum Structures and Quantum Control of Ministry of Education, Key Laboratory for Matter Microstructure and Function of Hunan Province, Department of Physics and Synergetic Innovation Center for Quantum Effects and Applications, Hunan Normal University, Changsha 410081, China}
\author{Jie-Qiao Liao}
\email{Corresponding author: jqliao@hunnu.edu.cn}
\affiliation{Key Laboratory of Low-Dimensional Quantum Structures and Quantum Control of Ministry of Education, Key Laboratory for Matter Microstructure and Function of Hunan Province, Department of Physics and Synergetic Innovation Center for Quantum Effects and Applications, Hunan Normal University, Changsha 410081, China}
\author{Hui Jing}
\email{Corresponding author: jinghui73@gmail.com}
\affiliation{Key Laboratory of Low-Dimensional Quantum Structures and Quantum Control of Ministry of Education, Key Laboratory for Matter Microstructure and Function of Hunan Province, Department of Physics and Synergetic Innovation Center for Quantum Effects and Applications, Hunan Normal University, Changsha 410081, China}
\author{Le-Man Kuang}
\email{Corresponding author: lmkuang@hunnu.edu.cn}
\affiliation{Key Laboratory of Low-Dimensional Quantum Structures and Quantum Control of Ministry of Education, Key Laboratory for Matter Microstructure and Function of Hunan Province, Department of Physics and Synergetic Innovation Center for Quantum Effects and Applications, Hunan Normal University, Changsha 410081, China}

\date{\today}

\begin{abstract}
Phase transition from the over-damping to under-damping states is a ubiquitous phenomenon in physical systems.
However, what kind of symmetry is broken associated with this phase transition remains unclear.
Here, we discover that this phase transition is determined by an anti-parity-time (anti-$\mathcal{PT}$) symmetry hidden in a single damping linear resonator, which is significantly different from the conventional anti-$\mathcal{PT}$-symmetric systems with two or more modes.
We show that the breaking of the anti-$\mathcal{PT}$ symmetry yields the phase transition from the over-damping to under-damping states, with an exceptional point (EP) corresponding to the critical-damping state.
Moreover, we propose an optomechanical scheme to show this anti-$\mathcal{PT}$ symmetry breaking by using the optical spring effect in a quadratic optomechanical system. We also suggest an optomechanical sensor with the sensitivity enhanced significantly around the EPs for the anti-$\mathcal{PT}$ symmetry breaking.
Our work unveils the anti-$\mathcal{PT}$ symmetry hidden in damping oscillations and hence opens up new possibilities for exploiting wide anti-$\mathcal{PT}$ symmetry applications in single damping linear resonators.
\end{abstract}

\maketitle

\section{Introduction}\label{introduction}

Damping oscillation is one of the most fundamental and important physical processes~\cite{Waves2014}, and such behavior appears in various physical systems, such as electronic, atomic, mechanical (acoustic), and optical (photonic) resonators.
It is well known that a damping resonator undergoes a phase transition from the monotonically damping without oscillation (over-damping) to an oscillatory behavior with a damping amplitude (under-damping), crossing the critical condition that the damping rate is twice the value of the resonant frequency.
However, there is no discussion on \emph{whether there exists some kind of symmetry breaking determining this phase transition.}
According to the quantum theory of open systems~\cite{Breuer2007}, the damping effect can be described simply with a damping term by eliminating the reservoir coupled to the resonator, and hence the symmetry in the damping oscillator should be discussed based on a non-Hermitian Hamiltonian including the damping term.
In the past decades, the non-Hermitian physics with parity-time ($\mathcal{PT}$) symmetry~\cite{BenderPRL98,BenderRPP07,Konotop2016RMP,El-Ganainy2018NatPh,WangHF2021JOpt} has attracted intense interest from photonics~\cite{GuoAPRL09,Christian2010NatPh,LinZPRL11,JingPRL14,LvXYPRL15,Peng014NatPh,Chang2014NaPho,OzdemirNM2019,MiriSCI19,Klauck2019NaPho,PengB2014Sci,Feng2014Sci,LiuZPPRL16,ChenWNature17,HodaeiNat17,WuQCSCPMA23} to acoustics~\cite{Bender2013AmJPh,ZhuXFPRX14,FleuryNC15,DingKPRX16,LiuTPRL18} and other communities~\cite{Schindler2011PRA,SchindlerJPA2012,YangXPRL2022,SunY2014PRL,Assawaworrarit2017Nat,DongZNE19,LiuHL2019SciA,WangWCPRA21,DingLYPRL21,WuY2019Sci,XiaoXSCPMA2020,ZhaoN2023PRL}.
However, a damping linear resonator does not satisfy the conditions of $\mathcal{PT}$ symmetry for the requirement of a gain to balance the loss.

Anti-$\mathcal{PT}$ symmetry~\cite{LiGPRA13} is another non-Hermitian symmetry attracted considerable theoretical and experimental interests~\cite{WuJHPRL14,HeY2022PRA,XuHTNat2016,Doppler2016Natur,WangYXPRA19,ZhangF2020PRL,Park2021PRL,ZhangNanoLett20,WuJHPRA15,PengP2016NatPh,JiangYPRL19,Choi2018NatCo,LiY2019Sci,YangF2017PRA,ZhangLight2019LSA,LiQOpt2019Optic,LaiYH2019Natur,PengZH2020PRA,Bergman2021NatCo,YangY2020PRL,ZhaoJPRAPP20,Zhang2021arXiv,ZhangHL2022CPB,ZhangXHPRA21,LuoXW2022PRL,Nair2021PRL}.
The breaking of anti-$\mathcal{PT}$ symmetry leads to exotic functionalities with pure dissipation, such as unidirectional reflectionless~\cite{WuJHPRL14,HeY2022PRA}, chiral mode switching~\cite{XuHTNat2016,Doppler2016Natur,WangYXPRA19}, and precision quantum sensing~\cite{ZhangNanoLett20,LuoXW2022PRL}.
A tremendous effort has also been devoted to achieving
anti-$\mathcal{PT}$ symmetry by using either dissipative coupling~\cite{WuJHPRA15,PengP2016NatPh,YangF2017PRA,JiangYPRL19,Choi2018NatCo,LiY2019Sci,ZhangLight2019LSA,LiQOpt2019Optic,LaiYH2019Natur,YangY2020PRL,PengZH2020PRA,Bergman2021NatCo,ZhaoJPRAPP20,Zhang2021arXiv,ZhangHL2022CPB} or parametric (nonlinear) driving~\cite{WangYXPRA19,ZhangF2020PRL,Park2021PRL}.
Nevertheless, the relation between the anti-$\mathcal{PT}$ symmetry breaking and phase transition from the over-damping to under-damping states has not been revealed yet.

In this work, we find that \emph{the dynamical Hamiltonian matrix of a damping linear resonator is anti-$\mathcal{PT}$ symmetric}, and that the breaking of the anti-$\mathcal{PT}$ symmetry induces the phase transition from the over-damping to under-damping states crossing the critical-damping state corresponding to an exceptional point (EP).
Different from the conventional anti-$\mathcal{PT}$-symmetric systems that two or more modes are required~\cite{WuJHPRA15,PengP2016NatPh,YangF2017PRA,JiangYPRL19,Choi2018NatCo,LiY2019Sci,ZhangLight2019LSA,LiQOpt2019Optic,LaiYH2019Natur,YangY2020PRL,PengZH2020PRA,Bergman2021NatCo,ZhaoJPRAPP20,Zhang2021arXiv}, here the anti-$\mathcal{PT}$ symmetry is found in a single-mode system, which greatly expands its range of applications.
Moreover, we propose an optomechanical scheme~\cite{Aspelmeyer2014RMP} to observe this anti-$\mathcal{PT}$ symmetry breaking for a damping mechanical resonator in a quadratic optomechanical system~\cite{ThompsonNat08,Sankey2010NatPh,Karuza2012JOP,Fonseca2016PRL,ShengJT2020PRL,Uros2019PRL,Bullier2021PRR,Purdy2010PRL,Hill2013PHD,Doolin2014PRA,Brawley2016NatCo,Kaviani2015Optica,TaofiqPRX15} based on the optical string effect~\cite{Virgilio2006PRA,Hossein2007OptL,TianF2014ApPhL}.
We show that the breaking of the anti-$\mathcal{PT}$ symmetry in a damping mechanical resonator can be used for ultra-sensitive sensing based on the frequency splitting around the EPs. The results presented in this work are general and hence these physics can be observed in other damping linear resonant platforms, including atomic~\cite{WuJHPRA15,PengP2016NatPh,JiangYPRL19}, electrical~\cite{Choi2018NatCo}, thermal~\cite{LiY2019Sci}, optical~\cite{YangF2017PRA,ZhangLight2019LSA,LiQOpt2019Optic,LaiYH2019Natur,PengZH2020PRA,Bergman2021NatCo}, magnonic~\cite{YangY2020PRL,ZhaoJPRAPP20} and mechanical~\cite{Zhang2021arXiv} systems.

\section{Anti-$\mathcal{PT}$ symmetry}

\begin{figure}[tbp]
\includegraphics[bb=89 190 491 717, width=8.5 cm, clip]{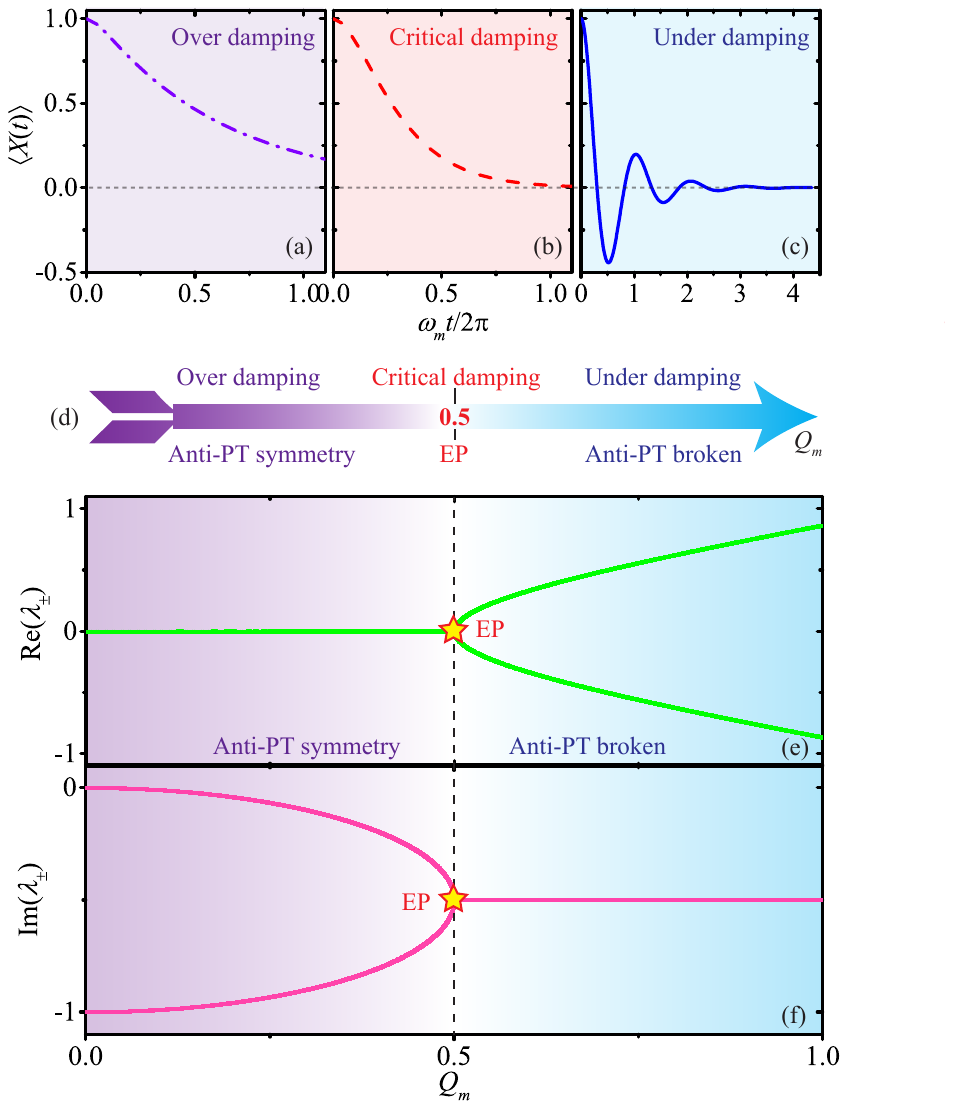}
\caption{(Color online) The dynamics of the mean value $\langle X \rangle \equiv \langle x \rangle/\sqrt{\hbar/(M\omega _{m})}$ for three damping states: (a) over damping ($Q_m=0.25$), (b) critical damping ($Q_m=0.5$), and (c) under damping ($Q_m=2$), under the initial conditions $\langle X(0)\rangle=1$ and $\langle p(0)\rangle=0$. (d) The correspondence between the dynamic states and the sorts of symmetry for a damping linear resonator. (e) Real and (f) imaginary parts of the eigenvalues $\lambda _{\pm}$ of $H_{\mathrm{dym}}/(\hbar \gamma _{m})$ with an EP at $Q_m=0.5$.}
\label{fig1}
\end{figure}

We consider a damping linear resonator with mass $M$ and spring constant $k$,
which is described by a Hamiltonian $H_{\rm MR}=p^2/(2M)+kx^2/2$ with the displacement $x$ and momentum $p$, and $\omega _{m}=\sqrt{k/M}$ is the resonant frequency of the mechanical resonator.
The equations of motion for the mean values $\langle x \rangle$ and $\langle p \rangle$ are given by
$d\langle x \rangle /dt= \langle p \rangle/M$ and $d\langle p \rangle/dt=-k \langle x \rangle -\gamma _{m} \langle p \rangle $,
where $\gamma _{m}$ is the damping rate induced by the couplings to the thermal reservoir~\cite{Bowen2015Book}.
We already know that the damping mechanical resonator exhibits over-damping, critical-damping, and under-damping states, corresponding to $\omega _{m}<\gamma _{m}/2$, $\omega _{m}=\gamma _{m}/2$, and $\omega _{m}>\gamma _{m}/2$, as shown in Figures~\ref{fig1}(a)-\ref{fig1}(c).
Although the phase transition for these dynamic behaviors is widely known, the associated intrinsic symmetry breaking has not been discovered yet.

Below we show that the dynamical Hamiltonian matrix of a damping mechanical resonator is anti-$\mathcal{PT}$ symmetric, and the over-, under- and critical-damping states are associated with the phases of anti-$\mathcal{PT}$ symmetry, anti-$\mathcal{PT}$ broken, and EP, respectively.
To reveal the anti-$\mathcal{PT}$ symmetry in a damping mechanical resonator, we rewrite the mechanical displacement and
momentum as $x=x_{\mathrm{zp}}\left( b^{\dag }+b\right)$ and $p=ip_{\mathrm{zp}}\left( b^{\dag }-b\right)$, with $x_{\mathrm{zp}}=\sqrt{\hbar /(2M\omega _{m})}$, $p_{\mathrm{zp}}=\sqrt{\hbar M\omega _{m}/2}$, and $[b, b^{\dag }]=1$. Then the equations of motion are rewritten as $i d \left (\langle b\rangle, \langle b^{\dag }\rangle\right)^{T}/dt =H_{\mathrm{dym}}\left( \langle b\rangle, \langle b^{\dag }\rangle\right)^{T}$ (``$T$" denoting matrix transpose), where the dynamical Hamiltonian matrix $H_{\mathrm{dym}}$ is given by
\begin{equation}\label{eq1}
\frac{H_{\mathrm{dym}}}{\hbar \gamma_{m}}=\left(
\begin{array}{cc}
 Q_{m}-i/2  & i/2
\\
i/2 &  -Q_{m}-i/2
\end{array}%
\right)
\end{equation}
with the quality factor $Q_{m}\equiv \omega_{m}/\gamma _{m}$.
The $H_{\mathrm{dym}}$ in Eq.~(\ref{eq1}) is anti-$\mathcal{PT}$-symmetric as
\begin{equation}
(\mathcal{PT})H_{\mathrm{dym}}(\mathcal{PT})^{-1}=-H_{\mathrm{dym}},
\end{equation}
with the parity operation $\mathcal{P}$ for switching $b \leftrightarrow b^{\dag }$ and the time-reversal operation $\mathcal{T}$ for complex conjugation.
The eigenvalues of $H_{\mathrm{dym}}/(\hbar \gamma _{m})$ are
\begin{equation}\label{eq2}
\lambda _{\pm}=-i/2 \pm \sqrt{ Q_{m}^{2}- 1/4},
\end{equation}
corresponding to the eigenstates $\Psi _{\pm }=(\beta _{\pm },\beta _{\pm }^{\prime
})^{T}$, with coefficients $\beta _{\pm }/\beta _{\pm }^{\prime }=-i 2 Q_m \pm \sqrt{1-4 Q_m ^{2}}$.
It should be mentioned that $\lambda _{\pm}$ in Eq.~(\ref{eq2}) are also the eigenvalues of the dynamical Hamiltonian matrix for the mean values of the dimensionless position $X\equiv x/(\sqrt{2}x_{\rm zp})$ and
momentum $P\equiv p/(\sqrt{2}p_{\rm zp})$, as $i d \left (\langle X\rangle, \langle P \rangle\right)^{T}/dt =H^{\prime}_{\mathrm{dym}}\left( \langle X\rangle, \langle P \rangle\right)^{T}$, with
\begin{equation}\label{Heffxp}
\frac{H'_{\mathrm{dym}}}{\hbar \gamma _{m}}=\left(
\begin{array}{cc}
0 & i Q_{m} \\
-i Q_{m} & -i%
\end{array}%
\right) .
\end{equation}%
So $\lambda _{\pm}$ in Eq.~(\ref{eq2}) are physical and can be measured experimentally.

The phases for the damping mechanical resonator can be discovered by analyzing both the real and imagine parts of the eigenvalues $\lambda _{\pm}$ [Figures~\ref{fig1}(e) and \ref{fig1}(f)]. We can see a phase transition from the anti-$\mathcal{PT}$-symmetric to the anti-$\mathcal{PT}$ broken phases occurring at $Q _{m}=0.5$.
In the anti-$\mathcal{PT}$-symmetric phase with $Q_{m}<0.5$, $\Psi _{\pm }$ are also the eigenstates of the parity-time operator $\mathcal{PT}$, corresponding to the over-damping state.
In the anti-$\mathcal{PT}$ broken phase with $Q_{m}>0.5$, $\Psi _{\pm }$ are no longer the eigenstates of the parity-time operator $\mathcal{PT}$, corresponding to the under-damping state.
Moreover, the critical point $Q _{m}=0.5$ for the anti-$\mathcal{PT}$ phase transition, namely EP, corresponds to the critical-damping state.
The correspondence between the dynamic states and the sorts of symmetry for a damping mechanical resonator is shown in Figure~\ref{fig1}(d).
As the symmetry of the damping resonator depends on the value of $Q_{m}$, the anti-$\mathcal{PT}$ symmetry breaking can be realized by adjusting either the decay rate $\gamma_m$ or the resonant frequency $\omega_{m}$ of the damping mechanical resonator.

\section{Optomechanical induced anti-$\mathcal{PT}$ symmetry breaking}

In order to observe anti-$\mathcal{PT}$ symmetry breaking in the single mechanical resonators, we propose to regulate the effective frequency of the mechanical resonator by the optical spring effect in this section.
Optomechanical systems possessing interactions between light and mechanical resonators provide an ideal platform to observe the mechanical anti-$\mathcal{PT}$ symmetry breaking~\cite{Aspelmeyer2014RMP}.
We consider a mechanical resonator coupled to an optical mode ($A$ and $A^{\dag }$, with frequency $\omega _{c}$) through quadratic optomechanical interaction, and the optical mode is driven resonantly by an external field with strength $\Omega $ and frequency $\omega _{L}=\omega _{c}$. In a
rotating frame with respect to $\hbar \omega _{L}A^{\dag }A$, the Hamiltonian of the system reads
\begin{equation}\label{Hom}
H_{\rm OM}=\frac{p^2}{2M}+\frac{1}{2}(k+ 2 g_0 A^{\dag }A)x^2+\hbar\Omega \left( A^{\dag }+A\right),
\end{equation}
where $ 2 g_0 A^{\dag }A$ is the spring constant induced by the optical mode with the single-photon quadratic optomechanical coupling strength $g_0$.

\begin{figure}[tbp]
\centering
\includegraphics[bb=92 249 487 689, width=8.5 cm, clip]{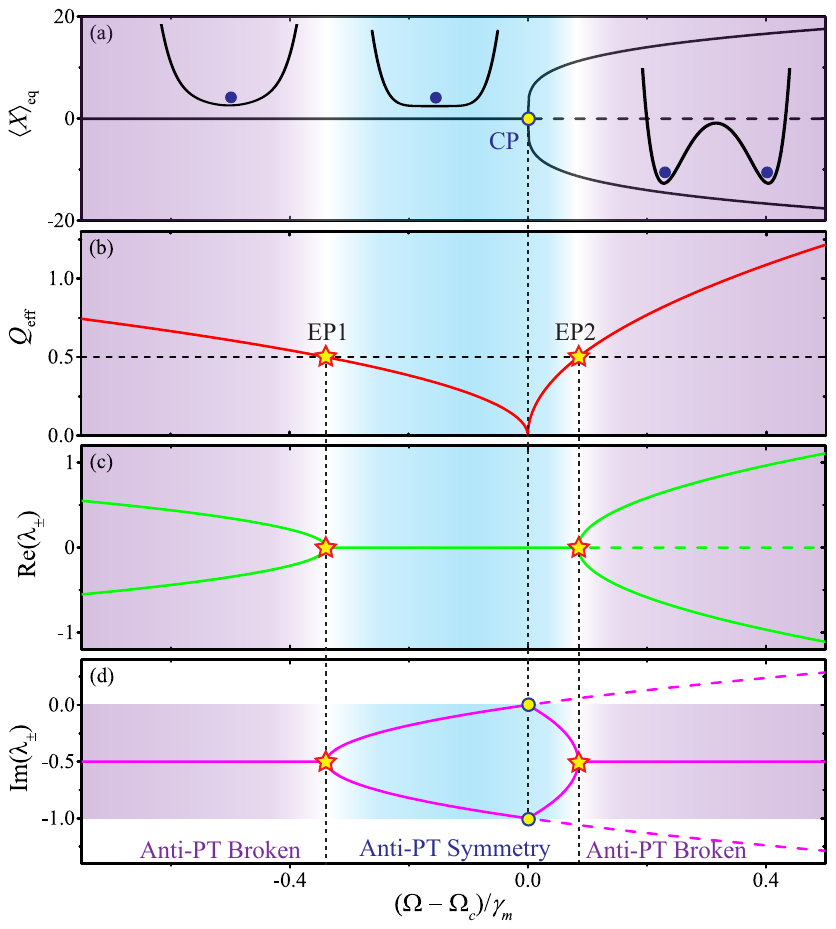}
\caption{(Color online) (a) Equilibrium positions of the dimensionless displacement $\langle X \rangle_{\rm eq}$ (effective potentials $U_{\mathrm{eff}}$ in the insets), (b) effective quality factor $Q_{\rm eff}$, (c) real parts of $\lambda_{\pm}$, and (d) imaginary parts of $\lambda_{\pm}$, versus the driving strength $\Omega$.}
\label{fig2}
\end{figure}

The quadratic optomechanical interaction has been demonstrated in various cavity-optomechanical systems, including
mechanical resonator (membrane~\cite{ThompsonNat08,Sankey2010NatPh,Karuza2012JOP,ShengJT2020PRL}, nanosphere~\cite{Fonseca2016PRL,Uros2019PRL,Bullier2021PRR}, cold atoms~\cite{Purdy2010PRL}) trapped in Fabry-Perot cavities,
or coupled to whispering-gallery-mode~\cite{Hill2013PHD,Doolin2014PRA,Brawley2016NatCo} or photonic crystal cavities~\cite{Kaviani2015Optica,TaofiqPRX15}.
For reality, we take the experimental parameters in our simulations for a planar silicon photonic crystal optomechanical cavity~\cite{TaofiqPRX15} with $g_0 =-6.06 \times 10^{-4}$ $\rm N/m$, $M=3.6$ pg, $\omega _{m}/2\pi=8.7$ MHz, $Q_{m}= 10^4$, and optical damping rate $\gamma _{c}/2\pi = 5$ GHz, i.e., corresponding to the sideband-unresolved regime $\gamma _{c}\gg \omega _{m}$.
The motions of the mechanical resonator can be measured by a weak probe beam bypass the cavity, and the power spectra of the weak probe beam exhibits the eigenfrequency of the mechanical resonator~\cite{ShengJT2020PRL}. But it should be pointed out that the negative frequency can not measured directly.

The dynamic behaviors of the mechanical resonator with quadratic optomechanical coupling can be analyzed based on the equilibrium positions obtained from the equations of motion: $d\langle x \rangle /dt= \langle p \rangle/M$ and $d\langle p \rangle/dt=-k' \langle x \rangle -\gamma _{m} \langle p \rangle $. Here, $k'=k+2g_{0}\Omega ^{2}/[( \gamma _{c}/2) ^{2}+(\left\langle x\right\rangle^{2} g_{0}/\hbar) ^{2}]$ is a position-dependent string constant obtained by adiabatically eliminating the optical mode in the sideband-unresolved regime (see \ref{appendix A} for details). By introducing an effective potential $U_{\mathrm{eff}} \equiv \int_{0}^{\left\langle x\right\rangle }k'\left\langle x^{\prime }\right\rangle d\left\langle x^{\prime }\right\rangle$, the equilibrium positions of the mechanical resonator are obtained under the conditions $dU_{\mathrm{eff}}/d\left\langle x\right\rangle =0\ $ and $d^{2}U_{\mathrm{eff}}/d\left\langle x\right\rangle ^{2}\geq 0$.
The equilibrium positions are shown in Figure~\ref{fig2}(a) with the effective potentials $U_{\mathrm{eff}}$ in the insets.
There is one equilibrium position at $\langle x\rangle _{\rm eq}=0$ under the critical driving $\Omega \leq \Omega _{c}=\sqrt{-k\gamma _{c}^{2}/(8g_{0})}$, and the bottom of potential becomes more and more flat (i.e., the spring softens) with the increasing driving strength $\Omega$.
When $\Omega > \Omega _{c}$, the position $\langle x \rangle=0$ becomes unstable, and the effective potential
is transformed into a double well with two new equilibrium positions at $\langle x\rangle _{\rm eq} = \pm (- \hbar /g_{0})^{1/2}[-(2g_{0}\Omega ^{2}/k)-(\gamma _{c}^2/4)]^{1/4}$.


To further analyze the motional feature of the mechanical resonator, we consider the small oscillation near its equilibrium positions $\langle x \rangle_{\rm eq}$. Thus, we have the effective spring constants $k_{\rm eff}=k(1-\Omega^2 /\Omega _{c}^2)$ for $\Omega \leq \Omega _{c}$ and $k_{\rm eff}=4k(1-\Omega_{c}^2 /\Omega^2)$ for $\Omega > \Omega _{c}$.
With the increase of the driving strength $\Omega$, the effective quality factor $Q_{\rm eff}\equiv \omega _{\rm eff}/\gamma_m$ with $\omega _{\rm eff}\equiv \sqrt{k_{\rm eff}/M}$ monotonously decreases for $\Omega \leq \Omega _{c}$, and then monotonically increases when $\Omega > \Omega _{c}$ [Figure~\ref{fig2}(b)].
The EPs can be demonstrated from the eigenvalues in Eq.~(\ref{eq2}) with $Q_m$ replaced by $Q _{{\rm eff}}$ [Figures~\ref{fig2}(c) and \ref{fig2}(d)].
There is one EP (EP1) under the critical driving $\Omega \leq \Omega _{c}$.
With the increase of $\Omega$, the EP1, corresponding to the transition from the anti-$\mathcal{PT}$-broken to anti-$\mathcal{PT}$-symmetry phases appears for $Q_{\rm eff}=0.5$ at the optical driving strength
\begin{equation}
\Omega _{\mathrm{EP1}}=\Omega _{c}\left(1-Q_m^{-2}/4\right)^{1/2}.
\end{equation}%
When the driving strength continues to increase ($\Omega > \Omega _{c}$), another EP (EP2), corresponding to the transition from the anti-$\mathcal{PT}$-symmetry to
anti-$\mathcal{PT}$-broken phases, appears at
\begin{equation}
\Omega _{\mathrm{EP2}}=\Omega _{c}\left(1-Q_m^{-2}/16\right)^{-1/2}
\end{equation}%
for $Q_{\rm eff}=0.5$.
It is worth noting that the conditions for the mechanical EPs are derived analytically by adiabatically eliminating the optical mode. In order to justify the validity of the method, we show that these results are in very good agreement with the prediction without making the adiabatical approximation (see \ref{appendix B} for more details).

The quadratic optomechanical system can not only be used to exhibit the anti-$\mathcal{PT}$ symmetry breaking, but also inspire a research for highly sensitive sensing based on the EPs.
The frequency bifurcating around the EP in the anti-$\mathcal{PT}$ broken phase provides a sensitive way to detect the small variations of the parameters caused by an external perturbation.

\section{Optomechanical anti-$\mathcal{PT}$ sensor}

High sensitivity is a long-term pursue goal due to the vital importance in both fundamental and applied physics.
As one of the most attractive points, the anti-$\mathcal{PT}$ symmetry breaking provides a method to enhance the sensitivity for sensing based on the frequency bifurcating around the EPs~\cite{ZhangNanoLett20,MaoX2020NJPh,LiT2021PRA,Djorwe2019PAAPP,ChenWNature17,LaiYH2019Natur}.
In the presence of a nanoparticle (scatterer) located around the optical cavity, one part of the optical mode field scatters into the environment, creating an addition damping, i.e., particle-induced scattering loss $\gamma_s$, which provides a physical mechanism for single nanoparticle detection~\cite{Xiao2017AdM}. Here, we consider a quadratic optomechanical system for sensing the particle-induced scattering loss $\gamma_s$ based on the frequency bifurcating around the EPs. We will show that the sensitivity can be enhanced sharply around the EPs.

\begin{figure}[tbp]
\centering
\includegraphics[bb=44 232 530 572, width=8.5 cm, clip]{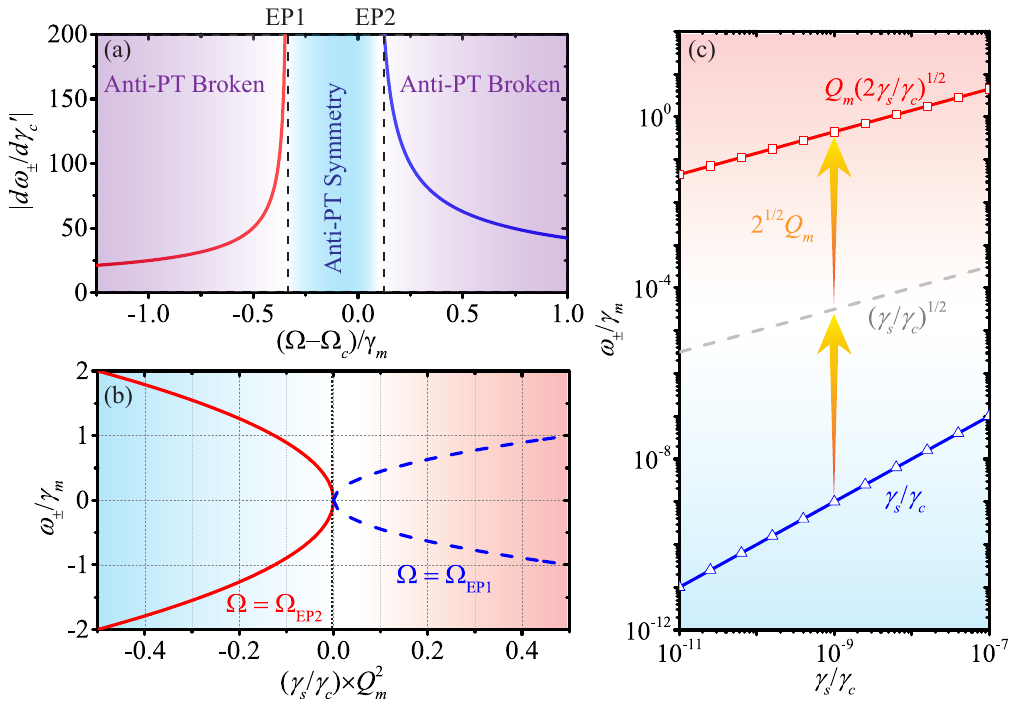}
\caption{(Color online) (a) The sensitivity $|d\omega_{\pm}/d\gamma^{\prime}_c|$ versus the driving strength $\Omega$. (b) Dependence of the frequency splittings $\omega_{\pm}$ on the scattering loss $\gamma_s$ at EP1 (blue dashed curve for $\Omega = \Omega _{\rm EP1}$) and EP2 (red solid curve for $\Omega = \Omega _{\rm EP2}$). (c) The comparison among $\omega_{\pm}=\sqrt{2}Q_{m}\sqrt{\gamma_s/\gamma _{c}}$, $\sqrt{\gamma_s/\gamma _{c}}$, and $\gamma_s/\gamma _{c}$.
}
\label{fig3}
\end{figure}

By considering the scattering loss $\gamma_s$ induced by a single nanoparticle, the total loss of the optical mode is given by $\gamma^{\prime}_c=\gamma_c+\gamma_s$. The sensitivity of the frequency splitting $\omega_{\pm }\equiv {\rm Re}(\lambda_{\pm }\gamma_m)$ on the particle-induced optical scattering loss $\gamma_s$ can be described by the derivative of $ \omega_{\pm }$ with respect to $\gamma^{\prime}_c$ as (see
\ref{appendix C} for details)
\begin{equation}
\frac{d\omega _{\pm }}{d\gamma^{\prime}_{c}}=\left\{
\begin{array}{ll}
\pm \frac{\omega _{m}^{2}}{\omega _{\pm }}\frac{\Omega ^{2}}{\Omega
_{c}^{\prime 2}}\frac{1}{\gamma ^{\prime}_{c}}, & \qquad \Omega <\Omega _{\mathrm{EP1}}, \\
0, & \qquad \Omega _{\mathrm{EP1}}<\Omega <\Omega _{\mathrm{EP2}}, \\
\mp \frac{4\omega _{m}^{2}}{\omega_{\pm }}\frac{\Omega_{c}^{\prime 2}}{\Omega
^{2}}\frac{1}{\gamma^{\prime}_{c}}, & \qquad  \Omega >\Omega _{\mathrm{EP2}},%
\end{array}%
\right. 
\end{equation}
where $\omega _{\pm }$ are shown in Eq.~(\ref{C1}) and $\Omega^{\prime} _{c}= \sqrt{-k\gamma _{c}^{\prime 2}/(8g_{0})}$.
The sensitivity is enhanced sharply as the driving strength going close to the EPs, and it becomes divergent as $|\omega _{\pm }|\rightarrow 0$ at the two EPs [Figure~\ref{fig2}(c)].
The sensitivities $|d\omega _{\pm }/d\gamma^{\prime}_{c}|$ around the EPs are shown in Figure~\ref{fig3}(a).

The dependence of the frequency splittings $\omega_{\pm }$ on the scattering loss $\gamma_s $ $(|\gamma_s| \ll \gamma _{c})$ is determined by the driving strength $\Omega$.
When $\Omega$ reaches the two EPs, the frequency splittings $\omega _{\pm }$ are approximately given by
\begin{equation}\label{omega_mp}
\frac{\omega _{\pm }}{\gamma _{m}}\approx \left\{
\begin{array}{ll}
\pm \sqrt{2}Q_{m}\sqrt{\frac{\gamma _{s}}{\gamma _{c}}}, & \qquad \Omega =\Omega _{%
\mathrm{EP1}},\; \gamma _{s}>0, \\
\pm 2\sqrt{2}Q_{m}\sqrt{\frac{-\gamma _{s}}{\gamma _{c}}}, & \qquad \Omega =\Omega
_{\mathrm{EP2}},\; \gamma _{s}<0.%
\end{array}%
\right. 
\end{equation}
We show the dependence of the frequency splittings $\omega _{\pm }$ on the scattering loss $\gamma_s$ in Figure~\ref{fig3}(b), corresponding to $\Omega =\Omega _{\mathrm{EP1}}$ (blue dashed curves) and $\Omega =\Omega _{\mathrm{EP2}}$ (red solid curves).
As $\omega _{\pm }/\gamma _{m}$ depend on the square root of $|\gamma_s|/\gamma_c$, the sensitivity can be enhanced sharply when $|\gamma_s|/\gamma_c \ll 1$.
Moreover, the factor $Q_{m}$ in Eq.~(\ref{omega_mp}) can further enhance the sensitivity significantly for the mechanical resonator in the cryogenic vacuum
conditions with a high quality factor $Q_m$ ranging from $10^4$ to $10^{7}$~\cite{ThompsonNat08,Sankey2010NatPh,Karuza2012JOP,Fonseca2016PRL,Uros2019PRL,ShengJT2020PRL,Bullier2021PRR,Purdy2010PRL,Hill2013PHD,Doolin2014PRA,Brawley2016NatCo,Kaviani2015Optica,TaofiqPRX15}.

To highlight the sensing enhancement based on EPs, we compare the frequency splittings $\omega _{\pm }$ corresponding to the driving strength around or far away from the EPs. The frequency splittings $\omega _{\pm }$ for the driving strength far away from the EPs become
\begin{equation}
\frac{\omega _{\pm }}{\gamma _{m}}\approx \left\{ 
\begin{array}{cc}
\pm Q_{\mathrm{eff}}\left( 1+\zeta \frac{\gamma _{s}}{\gamma _{c}}\right),  & \qquad
\Omega _{\mathrm{EP1}}-\Omega \gg \gamma _{m}, \\ 
\pm Q_{\mathrm{eff}}\left( 1-\zeta^{\prime} \frac{\gamma _{s}}{\gamma _{c}}\right),  & \qquad
\Omega -\Omega _{\mathrm{EP2}}\gg \gamma _{m},
\end{array}%
\right. 
\end{equation}
where $\zeta \equiv (\Omega ^{2}/\Omega _{c})^{2}/[ 1-(\Omega /\Omega _{c})^{2}]$, $\zeta^{\prime} \equiv (\Omega _{c} ^{2}/\Omega)^{2}/[ 1-(\Omega_{c} /\Omega )^{2}]$, and $Q_{\mathrm{eff}}$ depends on the driving strength $\Omega$.
There are two main differences in the expressions of the frequency splittings $\omega _{\pm }$ for the driving strength close to and far away from the EPs. Firstly, $ \omega _{\pm }$ depend on the square root of the scattering loss $\gamma_s$ for the driving strength close to the EPs, while $\omega _{\pm }$ depend linearly on the scattering loss $\gamma_s$ for the driving strength far away from the EPs. Secondly, there is an enhancement factor $Q_{m}$ in the expressions of $\omega _{\pm }$ for the driving strength close to the EPs, and $Q_{m}$ is replaced by $Q_{\mathrm{eff}}$ for the driving strength far away from the EPs.

To show the sensing enhancement more clearly, the comparison among $\omega_{\pm}=\sqrt{2}Q_{m}\sqrt{\gamma_s/\gamma _{c}}$, $\sqrt{\gamma_s/\gamma _{c}}$, and $\gamma_s/\gamma _{c}$ are shown in Figure~\ref{fig3}(c).
For $\gamma_s/\gamma _{c}=10^{-8}$, $\sqrt{\gamma_s/\gamma _{c}}$ are four orders of magnitude greater than $\gamma_s/\gamma _{c}$.
For $Q_{m}=10^{4}$, $\omega_{\pm}$ are enhanced by four orders of magnitude for $Q_{m}\sqrt{\gamma_s/\gamma _{c}}$ in comparing with $\sqrt{\gamma_s/\gamma _{c}}$.
Based on these two physical mechanisms, the ultra-sensitive sensing can be realized based on the frequency splittings around the EPs for the driving strength close to the EPs.
Similar mechanisms can be used to enhance the sensitivity for other optomechanical sensors~\cite{LI2013223PR,Liu2013OE,HeY2015ApPhL,YuW2016NatCo,LinQ2017PhRvA,LiuSP2019PRA,Sansa2020NatCo}.

\section{Conclusion}
In conclusion, we have revealed the hidden anti-$\mathcal{PT}$ symmetry breaking, which determines the phase transition from the over-damping to under-damping states in a damping linear resonator. We have also found that the critical-damping state corresponds to the EP.
In addition, we have proposed an optomechanical platform to exhibit this phase transition and exploited its application in ultra-sensitive sensing based on the frequency splittings around the EPs.
The physical principle between the anti-$\mathcal{PT}$ symmetry breaking and the damping-state phase transition is universal, and it governs the behaviors of all the damping linear resonant systems in physics,
mechanics, chemistry, and biology.
This work will open the door for exploring the wide applications~\cite{WuJHPRL14,HeY2022PRA,XuHTNat2016,Doppler2016Natur,WangYXPRA19,ZhangF2020PRL,Park2021PRL,ZhangNanoLett20,WuJHPRA15,PengP2016NatPh,YangF2017PRA,JiangYPRL19,Choi2018NatCo,LiY2019Sci,ZhangLight2019LSA,LiQOpt2019Optic,LaiYH2019Natur,YangY2020PRL,PengZH2020PRA,Bergman2021NatCo,ZhaoJPRAPP20,Zhang2021arXiv,ZhangXHPRA21,ZhangHL2022CPB,LuoXW2022PRL,Nair2021PRL} of anti-$\mathcal{PT}$ symmetry in single damping linear resonators,
and motivate the related studies on high-order EPs~\cite{Jing2017NatSR,LiuYL2017PRA,esee8c365,ZhangSM2020PRA,HuangR2022LPR} and non-Hermitian topological physics~\cite{Bergholtz2021RMP,HCWuPRB2021,ZHXuSCPMA2022,JSTang2022PRL}.

\emph{Note added.} After finishing this work, we note a related paper by Liang, Tang, Xu, and Liu~\cite{LiuYC2023PRL}, which has been published in Physical Review Letters recently. 
Liang \emph{et al.}~\cite{LiuYC2023PRL} proposed a scheme to realize a $\mathcal{PT}$-symmetry in a general dissipative resonance system by constructing a quadrature measurement-feedback loop,
and demonstrated the proposal experimentally in a thermal atomic ensemble.
Nevertheless, here we find that the dynamical Hamiltonian matrix of a damping linear resonator is anti-$\mathcal{PT}$ symmetric, without resort to a feedback loop, and show that the anti-$\mathcal{PT}$ symmetry breaking determines the phase transition from the over-damping to under-damping states.

\begin{acknowledgments}
This work was supported by the National Natural Science Foundation of China (Grant Nos. 12064010, 12247105, 12175061, 11935006, 11774086, 1217050862, and 11775075), the Natural Science Foundation of Hunan Province (Grant No. 2021JJ20036), and the science and technology innovation Program of Hunan Province (Grant Nos. 2022RC1203, 2020RC4047, 2021RC4029).
\end{acknowledgments}

\begin{appendix}

\section{Optomechanical induced mechanical anti-$\mathcal{PT}$ symmetry breaking}\label{appendix A}

In this section, we demonstrate the mechanical anti-$\mathcal{PT}$ symmetry breaking in a quadratic optomechanical system.
We introduce the quadratic optomechanical model and the quantum Langevin equations in Appendix~\ref{appendix A1}, derive the equilibrium positions of the mechanical resonator in Appendix~\ref{appendix A2}, and show the anti-$\mathcal{PT}$ symmetry breaking and the associated EPs in Appendix~\ref{appendix A3}.

\subsection{The quadratic optomechanical model and the quantum Langevin equations}\label{appendix A1}

We consider a mechanical resonator (with displacement $x$, momentum $p$, mass $M$, spring constant $k$, and resonant frequency $\omega _{m}=\sqrt{k/M}$) coupled to an optical mode (with annihilation operator $A$, creation operator $A^{\dag }$, and resonant frequency $\omega _{c}$) through a quadratical optomechanical interaction. The optical mode is driven by an external field with strength $\Omega$ and frequency $\omega _{L}$.
The Hamiltonian of the system reads
\begin{align}\label{Hsys1}
H_{\rm sys}=&\hbar \omega _{c}A^{\dag }A+\frac{p^{2}}{2M}+\frac{1}{2}k x^{2} + g_{0}A^{\dag }A x^{2} \nonumber\\
&+\hbar \Omega \left(e^{-i \omega _{L} t}A^{\dag }+e^{i \omega _{L} t} A\right),
\end{align}
where $g_{0}$ is the single-photon quadratic optomechanical coupling strength.
Below we consider the case where the optical mode is driven resonantly, i.e., $\omega _{L}=\omega _{c}$. Then in a rotating frame with respect to $\hbar \omega _{L}A^{\dag }A$, the Hamiltonian (\ref{Hsys1}) becomes
\begin{equation}
H_{\rm OM}=\frac{p^{2}}{2M}+%
\frac{1}{2}\left( k+2g_{0}A^{\dag }A\right) x^{2}+\hbar \Omega \left(
A^{\dag }+A\right).
\end{equation}%
In this work, we assume that the optical driving field is strong (i.e., $\Omega \gg
\gamma _{c}$, $\gamma _{c}$ is the damping rate of the optical
mode), and that the system works in the sideband-unresolved regime $\gamma _{c}\gg \omega _{m}$, which are consistent with the experimental parameters reported in the quadratic optomechanical systems~\cite{ThompsonNat08,Sankey2010NatPh,Karuza2012JOP,Fonseca2016PRL,Uros2019PRL,Bullier2021PRR,Purdy2010PRL,Hill2013PHD,Doolin2014PRA,Brawley2016NatCo,Kaviani2015Optica,TaofiqPRX15}.

By adding both the dissipation terms and the corresponding noise operators with zero expectation value into the Heisenberg equations of motion, we obtain the quantum Langevin equations for the system operators as
\begin{subequations}
\label{DM}
\begin{align}
\frac{dA}{dt}=&-\frac{\gamma _{c}}{2}A-i\frac{g_{0}}{\hbar }Ax^{2}-i\Omega +%
\sqrt{\gamma _{c}}A_{\rm in}, \label{DM1}\\
\frac{dx}{dt}=&\frac{p}{M},  \label{DM2}\\
\frac{dp}{dt}=&-\left( k+2g_{0}A^{\dag }A\right) x-\gamma _{m}p+F,\label{DM3}
\end{align}
\end{subequations}
where $A_{\rm in}$ is the optical input noise operator, $F$ is the stochastic force, and $\gamma _{m}$ is the mechanical damping rate. For the mechanical resonator, its quality factor is defined by $Q_{m}\equiv \omega_{m}/\gamma_{m}$.

\subsection{Equilibrium positions of the mechanical resonator}\label{appendix A2}

Before investigating the dynamic behaviors of the mechanical resonator, let us derive the equilibrium positions of the mechanical resonator, which can be obtained by analyzing the expectation values of the operators in the dynamical equations (\ref{DM}).
Based on the factorization assumptions $\langle Ax^{2} \rangle=\langle A\rangle\langle x\rangle^2$ and $\langle A^{\dag}Ax \rangle=|\langle A\rangle|^2 \langle x\rangle$~\cite{Aspelmeyer2014RMP,Bowen2015Book}, the expectation values of these operators satisfy the following classical equations of motion
\begin{subequations}
\label{DM_mean}
\begin{align}
\frac{d\left\langle A\right\rangle }{dt}=&-\frac{\gamma _{c}}{2}\left\langle
A\right\rangle -i\frac{g_{0}}{\hbar }\left\langle A\right\rangle
\left\langle x\right\rangle ^{2}-i\Omega, \label{eq.S13}\\
\frac{d\left\langle x\right\rangle }{dt}=&\frac{\left\langle p\right\rangle }{M}, \\
\frac{d\left\langle p\right\rangle }{dt}=& -\left( k+2g_{0}\left\vert
\left\langle A\right\rangle \right\vert ^{2}\right) \left\langle
x\right\rangle -\gamma _{m}\left\langle p\right\rangle. \label{eq.S15}
\end{align}
\end{subequations}
Below, we focus on the behaviors of the mechanical resonator by appropriately eliminating the optical mode.
In the sideband-unresolved regime in which the relaxation time of the optical mode is much shorter than the oscillating period of the mechanical resonator, the optical mode can be eliminated adiabatically by setting $d\left\langle A\right\rangle/dt=0$ in Eq.~(\ref{eq.S13}), then we obtain
\begin{equation}\label{Aexpress}
\left\langle A\right\rangle =\frac{-i\Omega }{\frac{\gamma _{c}}{2}+i\frac{%
g_{0}}{\hbar }\left\langle x\right\rangle ^{2}}.
\end{equation}
By substituting Eq.~(\ref{Aexpress}) into Eq.~(\ref{eq.S15}), the dynamic equation for the expectation value of the momentum can be obtained as
\begin{equation}
\frac{d\left\langle p\right\rangle }{dt}=-k'\left\langle
x\right\rangle -\gamma _{m}\left\langle p\right\rangle ,
\end{equation}%
where we introduce a position-dependent string constant
\begin{equation}
k'\equiv k+\frac{2g_{0}\Omega ^{2}}{\left( \frac{\gamma _{c}}{2}%
\right) ^{2}+\left( \frac{g_{0}}{\hbar }\left\langle x\right\rangle
^{2}\right) ^{2}}.
\end{equation}%
It is worth mentioning that the string constant $k'$ depends also on the external optical driving strength $\Omega$, which provides an approach to regulate the behaviors of the mechanical resonator on demand.

\begin{figure}[tbp]
\centering
\includegraphics[bb=64 222 546 662, width=6 cm, clip]{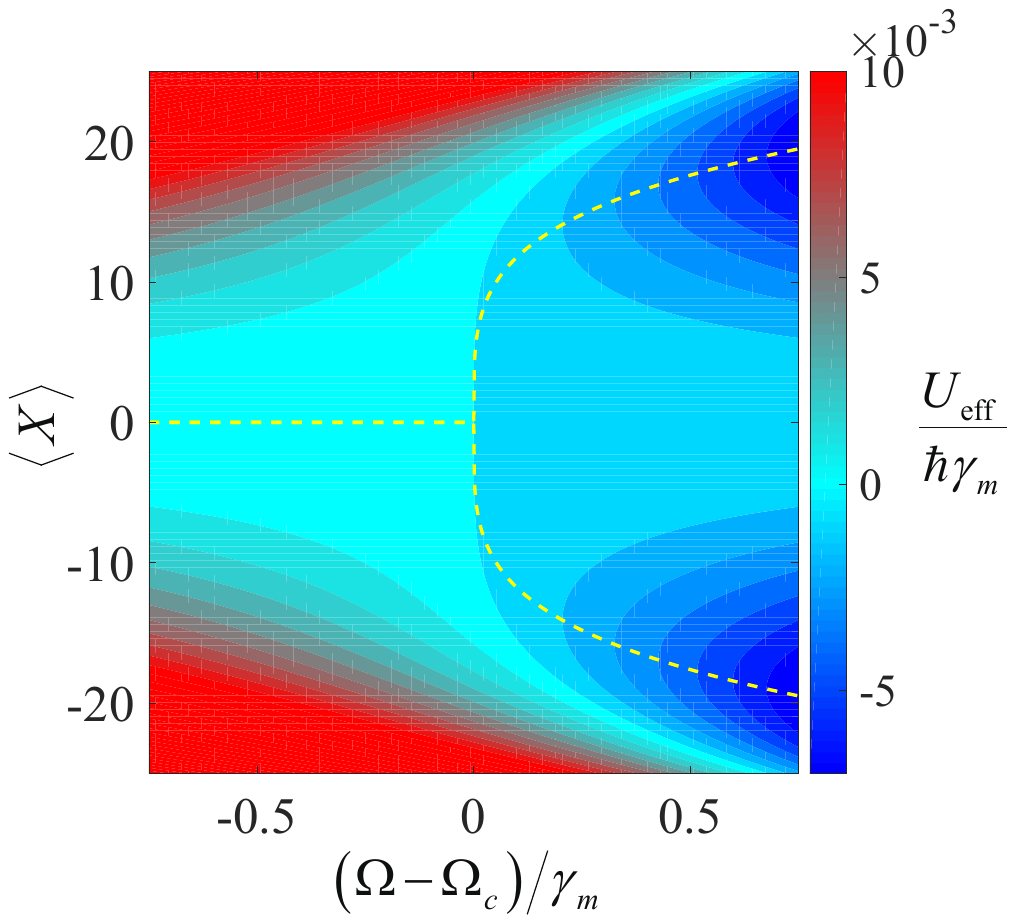}
\caption{(Color online) The effective potential $U_{\mathrm{eff}}/(\hbar \gamma_m)$ of the mechanical resonator versus both the driving strength $\Omega$ and the position $\langle X \rangle$. The equilibrium position $\langle X \rangle_{\rm eq}$ of the mechanical resonator is shown for reference (yellow dashed curves). Other parameters are: $g_0 =-6.06 \times 10^{-4}$ $\rm N/m$, $M=3.6$ pg, $\omega _{m}/2\pi=8.7$ MHz, $Q_{m} = 10^4$, and $\gamma _{c}/2\pi = 5$ GHz.}
\label{figA4}
\end{figure}

\begin{figure}[tbp]
\includegraphics[bb=29 346 527 550, width=8.5 cm, clip]{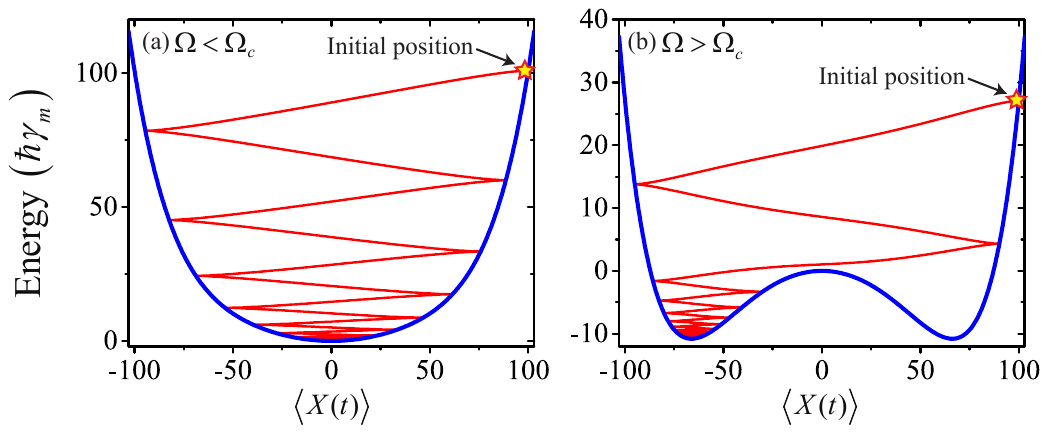}
\caption{(Color online) Relaxation of a mechanical resonator in (a) a single-well potential for $\Omega=\Omega_c-100 \gamma_m$ and (b) a double-well potential for $\Omega=\Omega_c+100 \gamma_m$.
The initial conditions are $\langle X(0) \rangle=100$, $\langle p(0) \rangle=0$, and $\left\langle A(0)\right\rangle =-i\Omega/[(\gamma _{c}/2)+i (g_{0}/\hbar)\langle x(0)\rangle ^{2}]$.
The blue curves are the effective potential $U_{\mathrm{eff}}/(\hbar \gamma_m)$ of the mechanical resonator given by Eq.~(\ref{Ueff}), and the red curves are the total energy of the mechanical resonator decaying with time.
Other parameters are the same as those used in Figure~\ref{figA4}.
}
\label{figA5}
\end{figure}

To further analyze the motional feature of the effective mechanical resonator, we seek for the effective potential to understand both the equilibrium position of the mechanical resonator and the action of the quadratic optomechanical interaction on the mechanical motion.
The effective potential is defined by $dU_{\mathrm{eff}}/d\left\langle x\right\rangle \equiv k'\left\langle
x\right\rangle $, which is obtained analytically as
\begin{equation}\label{Ueff}
U_{\mathrm{eff}} =\frac{1}{2}k\left\langle x\right\rangle ^{2}+\frac{2\hbar \Omega ^{2}}{%
\gamma _{c}}\arctan \left( \frac{2g_{0}\left\langle x\right\rangle ^{2}}{%
\hbar \gamma _{c}}\right)
\end{equation}
with the zero-potential energy position at $\left\langle x\right\rangle=0$.
To present an intuitive understanding, we show the effective potential $U_{\mathrm{eff}}$ of the mechanical resonator versus both the driving strength $\Omega$ and dimensionless position $\langle X\rangle\equiv \langle x\rangle/(\sqrt{2}x_{\rm zp})$ in Figure~\ref{figA4}, where $x_{\mathrm{zp}}=\sqrt{\hbar/(2M\omega _{m})}$ and the positions for the minimums of the potential energy (i.e., the equilibrium position $\langle X \rangle_{\rm eq}$ discussed in the following) are shown by yellow dashed curves.
In Figure~\ref{figA5}, we show the dynamics of the mechanical resonator based on the nonlinear equations of motion for the expectation values, i.e., Eq.~(\ref{DM_mean}). 
It is clear that the mechanical resonator oscillates with damping amplitude in a single-well potential for $\Omega < \Omega_c$ or in a double-well potential for $\Omega > \Omega_c$, which agree well with the effective potential given by Eq.~(\ref{Ueff}).

The equilibrium position $\left\langle x\right\rangle _{\rm eq}$ is obtained for the potential energy satisfying the conditions $dU_{\mathrm{eff}}/d\left\langle x\right\rangle =0 $ and $d^{2}U_{\mathrm{eff}}/d\left\langle x\right\rangle ^{2}\geq 0$.
From the condition $dU_{\mathrm{eff}}/d\left\langle x\right\rangle =0 $,
we have either
\begin{equation}
\left\langle x\right\rangle _{\rm eq}=0,
\end{equation}%
or%
\begin{equation}
\left\langle x\right\rangle _{\rm eq}=\pm \sqrt{- \frac{\hbar }{g_{0}}%
\sqrt{-\frac{2g_{0}\Omega ^{2}}{k}-\left( \frac{\gamma _{c}}{2}\right) ^{2}}}
\end{equation}
for $g_{0}<0$ and $\Omega >\Omega _{c}$.
Here, the critical driving strength is defined by
\begin{equation}\label{CP}
\Omega _{c}=\sqrt{-\frac{k\gamma _{c}^{2}}{8g_{0}}}.
\end{equation}%
From the condition $d^{2}U_{\mathrm{eff}}/d\left\langle x\right\rangle ^{2}\geq 0$, we have%
\begin{equation}
k+\frac{%
2g_{0}\Omega ^{2}}{\left( \frac{\gamma _{c}}{2}\right) ^{2}+\left( \frac{%
g_{0}}{\hbar }\left\langle x\right\rangle ^{2}\right) ^{2}}-\frac{%
8g_{0}\Omega ^{2}\left( \frac{g_{0}}{\hbar }\left\langle x\right\rangle
^{2}\right) ^{2}}{\left[ \left( \frac{\gamma _{c}}{2}\right) ^{2}+\left(
\frac{g_{0}}{\hbar }\left\langle x\right\rangle ^{2}\right) ^{2}\right] ^{2}}%
\geq 0.
\end{equation}%
Thus, $\left\langle x\right\rangle _{\rm eq}=0$ is stable only when the driving strength $\Omega \leq \Omega _{c}$, namely the single-well regime.
When the driving strength $\Omega >\Omega _{c}$, $\left\langle x\right\rangle =0$ becomes unstable (see Figure~\ref{figA4}), and there are two stable equilibrium positions at $\left\langle x\right\rangle _{\rm eq}=\pm \sqrt{-\frac{\hbar }{g_{0}}\sqrt{-\frac{2g_{0}\Omega ^{2}}{k}-\left( \frac{\gamma_{c}}{2}\right) ^{2}}}$, i.e., the double-well regime.

To sum up, we have the equilibrium positions
\begin{equation}
\left\langle x\right\rangle _{\rm eq}=\left\{
\begin{array}{cc}
0, & \qquad\Omega \leq \Omega _{c}, \\
\pm\sqrt{-\frac{\hbar }{g_{0}}\sqrt{-\frac{2g_{0}\Omega ^{2}}{k}-\left( \frac{%
\gamma _{c}}{2}\right) ^{2}}}, & \qquad\Omega >\Omega _{c}.%
\end{array}%
\right.
\end{equation}%
Meanwhile, the average photon number corresponding to the mechanical resonator in the equilibrium position read
\begin{equation}
 \left\vert \left\langle A\right\rangle_{\rm eq} \right\vert ^{2}=\left\{
\begin{array}{cc}
\frac{4\Omega ^{2}}{\gamma _{c}^{2}}, & \qquad \Omega \leq \Omega _{c}, \\
\frac{4\Omega _{c}^{2}}{\gamma _{c}^{2}}, & \qquad \Omega >\Omega _{c}.
\end{array}%
\right.
\end{equation}
Under the parameters in Figure~\ref{figA4}, we have the average photon number $\vert \langle A\rangle_{\rm eq}\vert ^{2} \approx 8875.6$ for $\Omega=\Omega_c$, which is not strong enough to consider the intrinsic nonlinear phenomena in the optical cavity~\cite{Leuthold2010NPho}.
We have checked that both the equilibrium positions $\left\langle x\right\rangle _{\rm eq}$ and the corresponding cavity coherent amplitude $\left\langle A\right\rangle_{\rm eq}$ are also the steady-state solutions of the equations of classical motion [Eqs.~(\ref{DM_mean})].

\subsection{Anti-$\mathcal{PT}$ symmetry breaking and the EPs}\label{appendix A3}

In this subsection, we analyze the motional feature of the mechanical resonator around the equilibrium position $\left\langle x\right\rangle _{\rm eq}$ to show the anti-$\mathcal{PT}$ symmetry breaking and the associated EPs.
The dynamical equations for $\langle x\rangle$ and $\langle p\rangle$ around the equilibrium position $\left\langle x\right\rangle _{\rm eq}$
can be obtained under the replacement: $\langle x\rangle \rightarrow \left\langle x\right\rangle _{\rm eq}+\langle x\rangle$ and $\langle p\rangle \rightarrow \langle p\rangle$ (for $\left\langle q\right\rangle _{\rm eq}=0$), as
\begin{subequations}
\begin{align}
\frac{d\left\langle x\right\rangle }{dt}=&\frac{\left\langle p\right\rangle }{M},\\
\frac{d\left\langle p\right\rangle }{dt}=&-k_{\mathrm{eff}}\left\langle
x\right\rangle -\gamma _{m}\left\langle p\right\rangle,
\end{align}
\end{subequations}
where we introduce the effective spring constant
\begin{equation}
k_{\mathrm{eff}}=\left\{
\begin{array}{lc}
k\left( 1-\frac{\Omega ^{2}}{\Omega _{c}^2} \right),  & \Omega \leq \Omega _{c}, \\
4k\left(1- \frac{\Omega_c ^{2}}{\Omega^{2}}\right) ,  & \Omega >\Omega _{c}
\end{array}%
\right.
\end{equation}
around the equilibrium positions. Based on the relations
\begin{equation}
x=x_{\mathrm{zp}}\left( b^{\dag }+b\right) , \quad p=ip_{\mathrm{zp}}\left( b^{\dag }-b\right)
\end{equation}
with $p_{\mathrm{zp}}=%
\sqrt{\hbar M\omega _{m}/2}$, the dynamical equations for $\langle x\rangle$ and $\langle p\rangle$ can also be rewritten as the dynamical equations for exception values of the bosonic annihilation operator $b$ and creation operator $b^{\dag }$ in a matrix form as
\begin{equation}
i\frac{d}{dt}\left(
\begin{array}{c}
\langle b\rangle \\
\langle b^{\dag } \rangle
\end{array}%
\right) =H_{\mathrm{eff}}\left(
\begin{array}{c}
\langle b \rangle \\
\langle b^{\dag }\rangle
\end{array}%
\right).
\end{equation}
Here, the dynamical Hamiltonian matrix for the effective mechanical resonator is given by%
\begin{equation}\label{Heff}
\frac{H_{\mathrm{eff}}}{\hbar \gamma _{m}}=\left(
\begin{array}{cc}
Q_{\mathrm{eff}}-i/2 & i/2 \\
i/2 & -Q_{\mathrm{eff}}-i/2%
\end{array}%
\right) ,
\end{equation}%
and the corresponding eigenvalues are given by
\begin{equation}\label{EgVpm}
\lambda _{\pm } =-i/2\pm \sqrt{Q_{\mathrm{eff}}^{2}-1/4},
\end{equation}
where $Q_{\mathrm{eff}}\equiv \omega_{\mathrm{eff}}/ \gamma _{m}$ is the quality factor of the effective mechanical resonator.
The $H_{\mathrm{eff}}$ [Eq.~(\ref{Heff})] is the same as the one for a simple mechanical resonator [Eq.~(1) in the main text], except that the quality factor $Q_{m}$ is replaced by the effective quality factor $Q_{\mathrm{eff}}$.
Importantly, the value of $Q _{\mathrm{eff}}$ depends on the driving strength $\Omega$, which provides an effective way to demonstrate both the mechanical anti-$\mathcal{PT}$ symmetry breaking and the EPs.

The driving strengths for the EPs can be given analytically for the system working around the equilibrium position $\left\langle x\right\rangle _{\rm eq}$.
In the single-well regime ($\Omega \leq \Omega_c$), we have
$\left\langle x\right\rangle _{\rm eq}=0$ and $ \omega _{\mathrm{eff}}=\sqrt{k_{\mathrm{eff}}/M} $ with $k_{\mathrm{eff}}=k\left( 1-\frac{\Omega ^{2}}{\Omega _{c}^2} \right)$.
There is one EP (EP1) at $Q_{\mathrm{eff}}=0.5$, and we can get the driving strength for EP1 as
\begin{equation}\label{EP1}
\Omega _{\mathrm{EP1}}=\Omega _{c}\left[ 1-\left( \frac{1}{%
2Q_{m}}\right) ^{2}\right]^{1/2}.
\end{equation}%
In the double-well regime ($\Omega > \Omega_c$), we have $\left\langle x\right\rangle _{\rm eq}=\pm \sqrt{-\frac{\hbar }{g_{0}}\sqrt{-\frac{2g_{0}\Omega ^{2}}{k}-\left( \frac{\gamma
_{c}}{2}\right) ^{2}}}$,
then another EP (EP2) appears at the driving strength
\begin{equation}\label{EP2}
\Omega _{\mathrm{EP2}}=\Omega _{c}\left[ 1-\left( \frac{1}{%
4Q _{m}}\right) ^{2}\right] ^{-1/2}.
\end{equation}

\section{The eigenvalues of the linearized Hamiltonian without adiabatical approximation}\label{appendix B}

In Appendix~\ref{appendix A1}, the mechanical EPs are derived analytically by adiabatically eliminating the optical mode based on the fact that the system works in the sideband-unresolved regime
$\gamma _{c}\gg \omega _{m}$.
In order to justify the validity of the method, here we calculate the eigenvalues of the linearized Hamiltonian for the quadratic optomechanical system without making the adiabatical approximation.

We expand the operators as the sum of the expectation values in the steady state (i.e., in the equilibrium position) and the quantum
fluctuations: $A\rightarrow \left\langle A\right\rangle_{\rm eq} +a$, $x\rightarrow
\left\langle x\right\rangle_{\rm eq} +x$, and $p\rightarrow p$, where $a$, $x$, and $p$ on the right-hand side of these relations are the quantum fluctuation operators, which satisfy the quantum Langevin equations
\begin{subequations}
\begin{align}
\frac{da}{dt}=&\left( -\frac{\gamma _{c}}{2}- i \Delta_{\rm eq}\right) a-iG_{\rm 0,eq}
x+\sqrt{\gamma _{c}}A_{\mathrm{in}}, \\
\frac{da^{\dag }}{dt}=&\left(-\frac{\gamma _{c}}{2}+i\Delta_{\rm eq}\right)a^{\dag }+iG^{*}_{\rm 0,eq}x+\sqrt{\gamma _{c}}A_{%
\mathrm{in}}^{\dag },  \\
\frac{dx}{dt}=&\frac{p}{M},   \\
\frac{dp}{dt}=&-k_{\rm eq} x -\hbar G_{\rm 0,eq} a^{\dag } -\hbar G^{*}_{\rm 0,eq} a -\gamma _{m}p+F.
\end{align}
\end{subequations}
with $\Delta_{\rm eq} \equiv \frac{g_{0}}{\hbar }\left\langle x\right\rangle_{\rm eq} ^{2}$,
$G_{\rm 0,eq}\equiv 2\frac{g_{0}}{\hbar }\left\langle A\right\rangle_{\rm eq}\left\langle x\right\rangle_{\rm eq} $,
and $k_{\rm eq}\equiv k+ 2g_{0}\left\vert \left\langle A\right\rangle_{\rm eq} \right\vert ^{2}$.
The equations can be rewritten with the dimensionless position $X\equiv x/(\sqrt{2}x_{\rm zp})$ and
momentum $P\equiv p/(\sqrt{2}p_{\rm zp})$ as
\begin{subequations}
\label{DM_qua}
\begin{align}
\frac{da}{dt}=&\left( -\frac{\gamma _{c}}{2}-i\Delta_{\rm eq}\right) a-iG_{\rm eq} X+%
\sqrt{\gamma _{c}}A_{\mathrm{in}},\label{eq.S44}\\
\frac{da^{\dag }}{dt}=&\left( -\frac{\gamma _{c}}{2}+i\Delta_{\rm eq}\right) a^{\dag }+i G_{\rm eq}^{\ast} X+\sqrt{\gamma _{c}}A_{\mathrm{in}}^{\dag },\\
\frac{dX}{dt}=&\omega _{m}P,\\
\frac{dP}{dt}=&-\omega'_{m} X-G_{\rm eq}^{\ast} a -G_{\rm eq} a^{\dag }-\gamma _{m}P+\xi,  \label{eq.S47}
\end{align}
\end{subequations}
with $\langle X\rangle_{\rm eq} \equiv \langle x\rangle_{\rm eq}/(\sqrt{2}x_{\rm zp})$, $g \equiv g_{0}/(2M\omega _{m})$, $\Delta_{\rm eq} \equiv 2g\left\langle X\right\rangle_{\rm eq}^{2}$,
$G_{\rm eq}\equiv 4g\left\langle A\right\rangle_{\rm eq} \left\langle X\right\rangle_{\rm eq}$,
$\omega'_{m}\equiv \omega _{m}+4g\left\vert \left\langle A\right\rangle_{\rm eq} \right\vert ^{2}$, and $\xi \equiv \sqrt{2}x_{%
\mathrm{zp}} F/\hbar$. By defining the vectors $V\equiv \left(
\begin{array}{cccc}
a, & a^{\dag }, & X, & P%
\end{array}%
\right) ^{T}$ and $V_{\mathrm{in}}\equiv \left(
\begin{array}{cccc}
\sqrt{\gamma _{c}}A_{in}, & \sqrt{\gamma _{c}}A_{in}^{\dag }, & 0, & \xi%
\end{array}%
\right) ^{T}$, the dynamical equations (\ref{DM_qua}) can be written in a matrix form as
\begin{equation}
\frac{d}{dt}V=-i H_{\mathrm{OM,eff}}V+V_{\mathrm{in}},
\end{equation}%
where we introduce the effective optomechanical Hamiltonian as
\begin{equation}\label{HOMeff}
H_{\mathrm{OM,eff}}= \left(
\begin{array}{cccc}
-i\frac{\gamma _{c}}{2}+\Delta_{\rm eq} & 0 & G_{\rm eq} & 0 \\
0 & -i\frac{\gamma _{c}}{2}-\Delta_{\rm eq} &
-G_{\rm eq}^{\ast} & 0 \\
0 & 0 & 0 & i\omega _{m} \\
-iG_{\rm eq}^{\ast} &
-iG_{\rm eq} & -i \omega'_{m}
& -i\gamma _{m}%
\end{array}%
\right) .
\end{equation}
The eigenvalues $\lambda_{i}$ ($i=1,2,3,4$) of $H_{\mathrm{OM,eff}}/\gamma _{m}$ are the roots of the characteristic polynomial $|H_{\mathrm{OM,eff}}/\gamma _{m} - \lambda I|=0$,
where $I$ is the identity matrix.

\begin{figure}[tbp]
\includegraphics[bb=92 325 457 586, width=8.5 cm, clip]{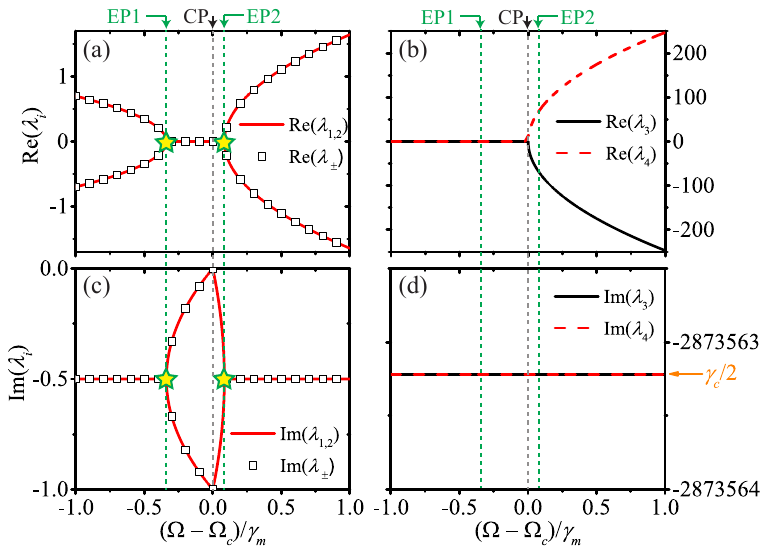}
\caption{(Color online) Comparison of these eigenvalues $\lambda_{i}$ ($i=1,2,3,4$) of $H_{\rm OM, eff}$ [Eq.~(\ref{HOMeff})] with the eigenvalues $\lambda_{\pm}$ of $H_{\rm eff}$ [Eq.~(\ref{Heff})] versus the driving strength $\Omega$: (a) and (b) real parts; (c) and (d) imaginary parts.
EP1 and EP2 are given by Eqs.~(\ref{EP1}) and (\ref{EP2}) and CP is given by Eq.~(\ref{CP}).
Other parameters are: $g_0 =-6.06 \times 10^{-4}$ $\rm N/m$, $M=3.6$ pg, $\omega _{m}/2\pi=8.7$ MHz, $Q_{m} = 10^4$, and $\gamma _{c}/2\pi = 5$ GHz.}
\label{figB6}
\end{figure}

In Figure~\ref{figB6}, we show $\lambda_{i}$ ($i=1,2,3,4$) and $\lambda_{\pm}$ [Eq.~(\ref{EgVpm})] versus the driving strength $\Omega$ for comparison. Two of $\lambda_{i}$ for $i=1,2$ are eigenvalues for the mechanical-like modes, and they are almost the same as $\lambda_{\pm}$ for both real and imaginary parts, as shown in Figures~\ref{figB6}(a) and \ref{figB6}(c). Here, the EPs can be confirmed with the prediction values given by Eqs.~(\ref{EP1}) and (\ref{EP2}).
Thus, the results in Eq.~(\ref{EgVpm}), which are obtained by eliminating the optical mode adiabatically, are justified.
Moreover, $\lambda_{i}$ ($i=3,4$) are the eigenvalues for the optical-like modes, with the frequency bifurcation around the critical driving strength $\Omega_c$, as denoted by CP (the abbreviation of ``critical point'') in Figures~\ref{figB6}(b) and \ref{figB6}(d).

\section{The sensitivity for optomechanical sensing based on EPs}\label{appendix C}

The EPs associated with the anti-$\mathcal{PT}$ symmetry breaking provide
a physical mechanism to enhance the sensing sensitivity in quantum sensing.
Here, based on the optomechanical induced mechanical anti-$\mathcal{PT}$ symmetry breaking, we consider ultrasensitive sensing on the scattering loss $\gamma_s$ induced by a single nanoparticle with the total loss of the optical mode given by $\gamma'_c=\gamma_c+\gamma_s$.
In terms of the quadratically optomechanical interaction and under the strong optical-driving condition, the frequency splittings $\omega _{\pm }$, i.e., the real part of the eigenvalues $\lambda _{\pm }\gamma_m$, are given by
\begin{equation} \label{C1}
\omega _{\pm } =\left\{
\begin{array}{ll}
\pm \sqrt{\omega _{m}^{2}\left( 1-\frac{\Omega ^{2}}{\Omega_{c}^{\prime 2}}\right)
-\left( \frac{\gamma _{m}}{2}\right) ^{2}}, & \: \Omega <\Omega _{\mathrm{EP1}},
\\
0, & \: \Omega _{\mathrm{EP1}}<\Omega <\Omega _{\mathrm{EP2}}, \\
\pm \sqrt{4\omega _{m}^{2}\left( 1-\frac{\Omega _{c}^{\prime 2}}{\Omega ^{2}}%
\right) -\left( \frac{\gamma _{m}}{2}\right) ^{2}}, & \: \Omega >\Omega _{%
\mathrm{EP2}}.%
\end{array}%
\right.
\end{equation}%
where $\Omega^{\prime} _{c}= \sqrt{-k\gamma _{c}^{\prime 2}/(8g_{0})}$.
The sensitivity of the frequency splittings $\omega _{\pm }$ on the change of the optical damping rate $\gamma^{\prime}_{c}$ can be described by the derivative of $\omega _{\pm }$ with respect to $\gamma^{\prime}_{c}$ as
\begin{equation}
\frac{d\omega _{\pm }}{d\gamma^{\prime } _{c}}=\left\{ 
\begin{array}{ll}
\pm \frac{\omega _{m}^{2}}{\omega _{\pm }}\frac{\Omega ^{2}}{\Omega
_{c}^{\prime 2}}\frac{1}{\gamma _{c}^{\prime }}, & \qquad \Omega <\Omega _{\mathrm{%
EP1}}, \\ 
0, & \qquad \Omega _{\mathrm{EP1}}<\Omega <\Omega _{\mathrm{EP2}}, \\ 
\mp \frac{4\omega _{m}^{2}}{\omega _{\pm }}\frac{\Omega _{c}^{\prime 2}}{%
\Omega ^{2}}\frac{1}{\gamma _{c}^{\prime }}, & \qquad \Omega >\Omega _{\mathrm{EP2}},%
\end{array}%
\right. 
\end{equation}
where $\omega _{\pm }$ are shown in Eq.~(\ref{C1}).

The dependence of the frequency splittings $\omega_{\pm }$ on the scattering loss $\gamma_s$ is determined by the driving strength $\Omega$.
Under the assumption that $\gamma_s\ll \gamma_c$, we have
\begin{equation}
\frac{\omega _{\pm }}{\gamma _{m}}\approx \left\{ 
\begin{array}{ll}
\pm \sqrt{Q_{\mathrm{eff}}^{2}- \frac{1}{4}+2Q_{m}^{2}\frac{\Omega ^{2}}{%
\Omega _{c}^{2}}\frac{\gamma _{s}}{\gamma _{c}}}, & \; \Omega <\Omega _{\mathrm{%
EP1}}, \\ 
0, & \; \Omega _{\mathrm{EP1}}<\Omega <\Omega _{\mathrm{EP2}}, \\ 
\pm \sqrt{Q_{\mathrm{eff}}^{2}-\frac{1}{4}-8 Q _{m}^{2}\frac{%
\Omega _{c}^{2}}{\Omega ^{2}}\frac{\gamma _{s}}{\gamma _{c}}}, & \; \Omega
>\Omega _{\mathrm{EP2}},%
\end{array}%
\right. 
\end{equation}
where $Q_{\mathrm{eff}}=Q_{m}\sqrt{1-(\Omega /\Omega_{c})^{2}}$ for $\Omega<\Omega _{\mathrm{EP1}}$ and $%
Q_{\mathrm{eff}}=2Q_{m}\sqrt{1-(\Omega _{c}/\Omega) ^{2}}$ for $\Omega >\Omega _{\mathrm{EP2}}$.
To show the sensing enhancement by the EPs, we compare the frequency splittings $\omega _{\pm }$ for the driving strength around the EPs with the results for the driving strength far away from the EPs.
We have $Q_{\mathrm{eff}}=0.5$ for both $\Omega =\Omega _{\mathrm{EP1}}$ and $\Omega =\Omega _{\mathrm{EP2}}$, so the frequency splittings $\omega _{\pm }$ around the two EPs are approximately given by
\begin{equation}
\frac{\omega _{\pm }}{\gamma _{m}}\approx \left\{ 
\begin{array}{ll}
\pm \sqrt{2} Q_{m}\sqrt{\frac{\gamma _{s}}{\gamma _{c}}}, & \qquad \Omega =\Omega _{%
\mathrm{EP1}},\; \gamma _{s}>0 ,\\ 
\pm 2\sqrt{2}Q_{m}\sqrt{\frac{-\gamma _{s}}{\gamma _{c}}}, & \qquad \Omega =\Omega _{%
\mathrm{EP2}},\; \gamma _{s}<0,
\end{array}%
\right. 
\end{equation}
for $\Omega _{\mathrm{EP1}}/\Omega_{c} \approx \Omega_{c}/\Omega _{\mathrm{EP2}} \approx 1$.
Instead, we have $Q_{\mathrm{eff}} \gg 0.5$ for $\left( \Omega _{\mathrm{EP1}}-\Omega \right) \gg \gamma _{m}$ or $\left( \Omega -\Omega _{\mathrm{EP2}}\right) \gg \gamma _{m}$,
then the frequency splittings $\omega _{\pm }$ corresponding to the driving strength $\Omega$ far
away from the two EPs are shown as 
\begin{equation}
\frac{\omega _{\pm }}{\gamma _{m}}\approx \left\{ 
\begin{array}{cc}
\pm Q_{\mathrm{eff}}\left( 1+\zeta \frac{\gamma _{s}}{\gamma _{c}}\right),  & \qquad
\Omega _{\mathrm{EP1}}-\Omega \gg \gamma _{m}, \\ 
\pm Q_{\mathrm{eff}}\left( 1-\zeta^{\prime} \frac{\gamma _{s}}{\gamma _{c}}\right),  & \qquad
\Omega -\Omega _{\mathrm{EP2}}\gg \gamma _{m},
\end{array}%
\right. 
\end{equation}
where $\zeta \equiv (\Omega ^{2}/\Omega _{c})^{2}/[ 1-(\Omega /\Omega _{c})^{2}]$ and $\zeta^{\prime} \equiv (\Omega _{c} ^{2}/\Omega)^{2}/[ 1-(\Omega_{c} /\Omega )^{2}]$.

\end{appendix}

\end{document}